\documentclass[prx, letterpaper, superscriptaddress, amsmath, amssymb, reprint, floatfix, nofootinbib]{revtex4-2}

\usepackage{graphicx}%
\usepackage{multirow}%
\usepackage{amsthm}%
\usepackage{mathrsfs}%
\usepackage[title]{appendix}%
\usepackage[usenames]{xcolor}%
\usepackage{textcomp}%
\usepackage{booktabs}%
\usepackage{listings}%
\usepackage{verbatim}

\usepackage{dcolumn}
\usepackage{bm}
\usepackage{tabularx}
\usepackage{array}
\usepackage{upgreek}
\usepackage{makecell} 
\usepackage[normalem]{ulem}

\usepackage{shellesc}

\makeatletter
\newcommand*{\addFileDependency}[1]{
  \typeout{(#1)}
  \@addtofilelist{#1}
  \IfFileExists{#1}{}{\typeout{No file #1.}}
}
\makeatother

\newcommand{\eqr}[1]{Eq.\ (\ref{#1})}
\newcommand{\ket}[1]{\left\vert{#1}\right\rangle}
\newcommand{\bra}[1]{\left\langle{#1}\right\vert}
\newcommand{\braket}[1]{\left\langle{#1}\right\rangle}

\newcommand{\diagproj}[1]{\ket{#1}\bra{#1}}

\newcommand{\gten}{\mathbf{g}}

\definecolor{firebrick}{HTML}{B22222}

\newcommand{\polcol}[1]{\textcolor{blue}{#1}}
\newcommand{\ancol}[1]{\textcolor{red}{#1}}
\newcommand{\singcol}[1]{\textcolor{firebrick}{#1}}

\newcommand{\szt}{\ket{\singcol{S_{02}}}}

\newcommand{\upua}{\ket{\polcol{\uparrow_{\rm P}}, \ancol{\uparrow_{\rm A}}}}
\newcommand{\upda}{\ket{\polcol{\uparrow_{\rm P}}, \ancol{\downarrow_{\rm A}}}}
\newcommand{\dpua}{\ket{\polcol{\downarrow_{\rm P}}, \ancol{\uparrow_{\rm A}}}}
\newcommand{\dpda}{\ket{\polcol{\downarrow_{\rm P}}, \ancol{\downarrow_{\rm A}}}}

\begin{document}

\preprint{APS/123-QED}

\title{Polarimetry With Spins in the Solid State}

\author{Lorenzo Peri}
\email{lp586@cam.ac.uk}
\affiliation{Quantum Motion, 9 Sterling Way, London, N7 9HJ, United Kingdom}
\affiliation{Cavendish Laboratory, University of Cambridge, JJ Thomson Ave, Cambridge CB3 0HE, United Kingdom}

\author{Felix-Ekkehard von Horstig}

\affiliation{Quantum Motion, 9 Sterling Way, London, N7 9HJ, United Kingdom}
\affiliation{Department of Materials Sciences and Metallurgy, University of Cambridge, Charles Babbage Rd, Cambridge CB3 0FS, United Kingdom}
 
\author{Sylvain Barraud}
\affiliation{CEA, LETI, Minatec Campus, F-38054 Grenoble, France}
 
\author{Christopher J. B. Ford}
\affiliation{Cavendish Laboratory, University of Cambridge, JJ Thomson Ave, Cambridge CB3 0HE, United Kingdom}

\author{M\'onica Benito}
\affiliation{Institute of Physics, University of Augsburg, Augsburg, 86159, Germany}
 
\author{M. Fernando Gonzalez-Zalba}
\affiliation{Quantum Motion, 9 Sterling Way, London, N7 9HJ, United Kingdom}

\date{\today}

\begin{abstract}

    The ability for optically active media to rotate the polarization of light is the basis of polarimetry, an illustrious technique responsible for many breakthroughs in fields as varied as astronomy, medicine and material science. Here, we recast the primary mechanism for spin readout in semiconductor-based quantum computers, Pauli spin-blockade (PSB), as the natural extension of polarimetry to the third dimension. We perform polarimetry with spins through a silicon quantum dot exchanging a hole with a boron acceptor, demonstrating the role of spin-orbit coupling in creating spin misalignment. Perfect spin alignment may be recovered by means of rotating the applied magnetic-field orientation.
    This work shows how spin misalignment sets a fundamental upper limit for the spin readout fidelity in quantum-computing systems based on PSB.

\end{abstract}

\maketitle


\section*{Introduction} 

Ever since its introduction in 1834 by William Fox Talbot \cite{Talbot_1834,Talbot_1835,Flugge_1900}, polarimetry has been one of the main techniques for investigating the world around us \cite{Ronchi_1970}. 
From tissues \cite{Ghosh_Vitkin_2011} to stars \cite{Tinbergen_2005}, from radars and navigation \cite{Kostinski_Boerner_1986,Li_2023,Kristjansson_2002,Ropars_2011,Valery_2022} to meteorology and industrial processes \cite{Brosseau_1998,Bates_1942}, from elucidating the nature of electromagnetic waves \cite{Pedrotti_2018} to understanding the behavior of chemical bonds \cite{Hegstrom_Kondepudi_1990}, polarimetry has provided an unprecedented eye into the microscopic properties of matter and their interaction with a rotationally quantized property, such as the polarization of light \cite{Gross_2012,Delly_2019}.

\begin{figure*}[htb!]
    \centering
    \includegraphics[width = \textwidth]{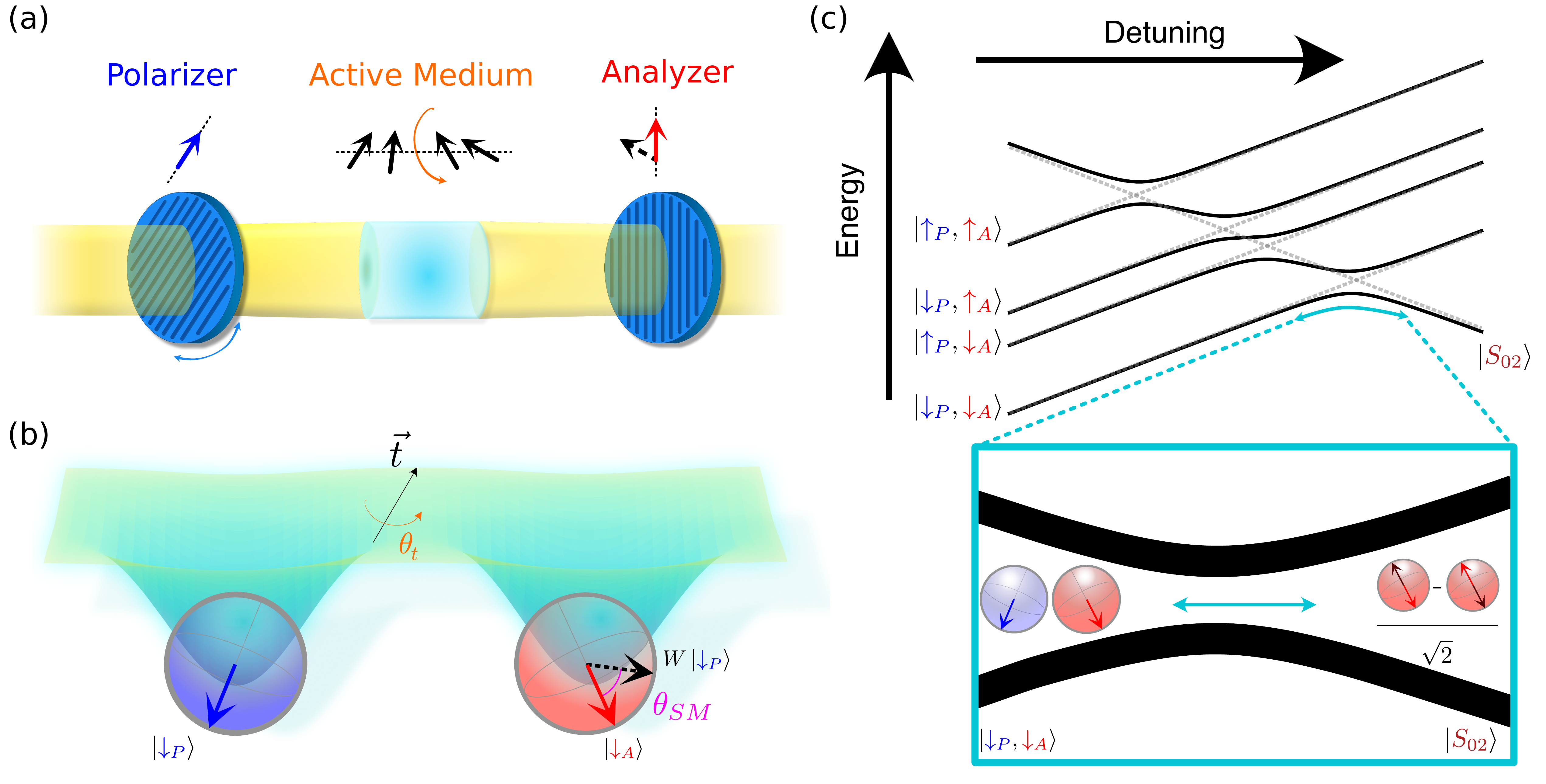}
    \caption{\textbf{Polarimetry with light and with spins.} (a) A standard light polarimeter. An active medium (capable of rotating the polarization axis) is placed between two polarizing elements (\textit{polarizer} and \textit{analyzer}) and the polarization misalignment angle is measured. (b) A spin-orbit-coupled double quantum dot seen as a polarimeter for spins. The two sites have different quantization axes, and spin-orbit coupling causes a spin rotation upon charge tunneling. (c) Energy diagram of an even-parity charge transition between $(1,1)$ and $(0,2)$ occupation. Spin-orbit coupling causes the crossing between the $(1,1)$ triplet and the $(0,2)$ singlet to be avoided (inset), lifting Pauli spin blockade.}
    \label{fig1:polarimeter}
\end{figure*}

\textit{In essentia}, the idea of polarimetry stems from the observation that particular degrees of freedom are \textit{polarized} (i.e. quantized along an axis in orthogonal states), and that certain physical systems are \textit{active}, in the sense that they break the relevant symmetry and may connect such states \cite{Jerrard_1982}\footnote{A theory ``\textit{the wonderful simplicity of which is such as to bear with it the stamp of truth}''. G.G. Stokes, 1851, Ref. \cite{Stokes_1851}}. A breakthrough from this original view by George Gabriel Stokes \cite{Stokes_1851} was introduced by Jones and later by Mueller \cite{Collett_2012}, who posited that (optical) \textit{activity} is best represented as a \textit{rotation} of the polarization (i.e., quantization) axis, plus a potential additional phase \cite{Azzam_2016}.
This intuition provides the basis of the polarimeter (Fig.~\ref{fig1:polarimeter}(a)), in which an active medium is placed between two polarizing elements (one moveable, the \textit{polarizer}, and one fixed, the \textit{analyzer}). The optical transmission is measured as a function of the polarizer angle, which quantifies the polarization misalignment between the analyzer and the image of the polarizer after rotation by the medium.

In this work, we extend the concept of polarimetry to another physical system of technological interest that manifests a polarized degree of freedom: spins confined in semiconductor quantum dots (QDs). We demonstrate how this platform, a promising candidate for quantum information processing \cite{Loss_DiVincenzo_1998, Zalba_2021, Scappucci2021, Philips_2022, Burkard_2023}, represents the natural extension of light polarimetry to the third dimension, and how concepts borrowed from the field of polarimetry provide valuable insight into the complex physics of coupled spins in the solid state, particularly when subject to spin-orbit coupling (SOC).
Here, we exploit notions from polarimetry to provide a geometrical intuition for the effect of SOC on spin qubits, introducing the concept of the spin misalignment angle between two QDs (acting as analyzer and polarizer).
We experimentally demonstrate our model by performing polarimetry with spins in a silicon double quantum dot (DQD).
Exploring different orientations of the applied magnetic field (polarizer direction), we measure the resulting spin misalignment, which vanishes at one particular \textit{magic angle} \cite{Sen_2023,Danon_Nazarov_2009}, where the (rotated) image of the polarizer perfectly projects onto the analyzer.
This corresponds to the physical phenomenon of Pauli spin-blockade (PSB), and it is the direct equivalent of perfect transmission through a light polarimeter.

For the purpose of quantum computation, SOC is attracting growing interest \cite{Hendrickx_2021,Jirovec2021, vanRiggelen_2024,Zhang_2024, Wang2024sci} thanks to the enticing opportunity of all-electrical universal qubit control \cite{Mutter_Burkard_2021,Liles_2024,Froning_2021}. 
Yet, it inherently introduces additional challenges.
The presence of spin misalignment caused by SOC manifests itself as a lifting of the PSB, undermining the effectiveness of the spin-to-charge conversion (SCC) used by many spin-readout schemes \cite{Hendrickx_2021,Oakes_2023,Lundberg_2024,Danon_Nazarov_2009,Han_2023}.
The analogy with polarimetry naturally leads us to introduce the concept of SCC fidelity to quantify such errors in the SCC process.
This effect is notably missing from the spin-qubit literature, and mandates a necessary correction (and sets a fundamental upper limit) for the fidelity of PSB-based spin readout, which, given more than two readout sites, may globally be below the fault-tolerance threshold for error correction, posing serious challenges for the scalability of spin-qubit architectures for quantum information processing when subject to SOC.

\section*{Spin-Orbit Coupling as an Active Medium}

\begin{figure*}[htb!]
    \centering
    \includegraphics[width = \textwidth]{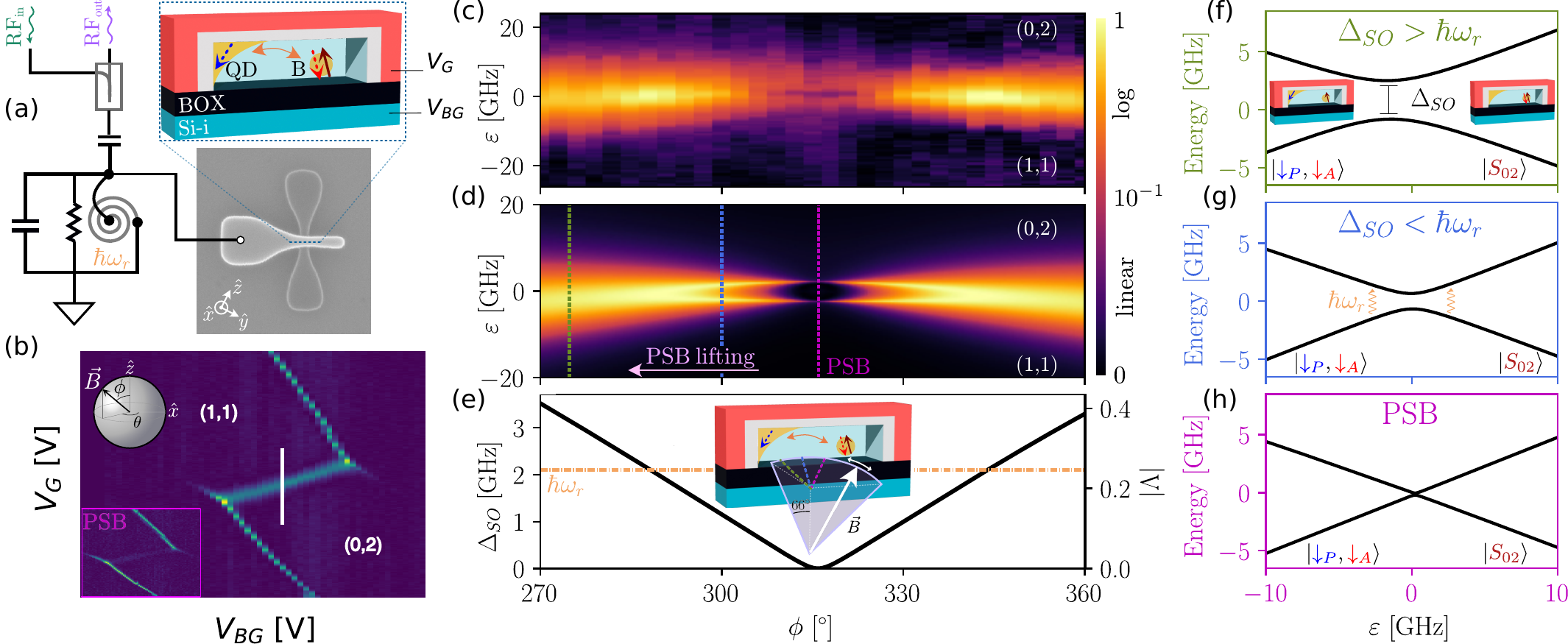}
    \caption{\textbf{A polarimeter for spins.} (a) A p-type silicon nanowire transistor hosting a spin-orbit-coupled double quantum dot (QD), composed of an electrostatically defined QD and a boron acceptor (B). Micrograph of the device coupled to a microwave resonator ($\omega_{\rm r}/2\pi=2.1$\,GHz), measured by reflectometry. The frame of reference for the magnetic field $\vec{B}$ is the same as for the vector magnet in the dilution refrigerator. (b) Charge-stability map measured at high applied external field ($|\vec{B}|=0.5$\,T) of the (nominal) $(1,1)\leftrightarrow(0,2)$ charge transition, in directions showing PSB (inset) and its lifting (main plot). (c-e) Dispersive sensing of the transition for changing magnetic-field angle (data in (c) and theory in (d)) showing PSB and resonant splitting, and (e) fitted spin-misalignment parameter $|\Lambda|$ (right axis) and spin-orbit gap ($\Delta_{\rm SO} = \sqrt{2} \tilde{t}_{\rm c} |\Lambda|$) (left axis). (f-h) Energy diagrams of the spin-orbit anticrossing for various spin alignments.}
    \label{fig2:phi}
\end{figure*}

The presence of SOC in a DQD manifests itself in two ways. Firstly, it causes two spatially separated spins to experience a different magnetic field, which is encoded in different \textit{g}-tensors $\gten$ for the two QDs. This effectively creates two separate (and generally misaligned) quantization directions $\gten\vec{B}$ (known as \textit{internal} fields) onto which each spin aligns when an \textit{external} magnetic field $\vec{B}$ is applied. 
However, SOC also allows inter-dot tunneling to mix spin states. Namely, tunneling may either \textit{conserve} or \textit{flip} the spin orientation.
The former process can be described by a scalar $t_0$, while the latter is not rotationally invariant and thus is most generally represented as a (real) vector $\vec{t}$ \cite{Danon_Nazarov_2009}. A key observation is that tunneling is not itself \textit{projective} (where the spin orientation would be \textit{either} conserved \textit{or} flipped). Thus, SOC is best understood as a tunneling process accompanied by a spin-axis \textit{rotation} around the $\vec{t}$ vector of an angle $\theta_{t}/2 = \arctan\left(|\vec{t}|/t_0\right)$, represented by the unitary transformation $W = \exp{\left(\frac{\theta_{t}}{2i} \frac{\vec{t}}{|\vec{t|}} \cdot \vec{\mathbf{\sigma}}\right)}$ \cite{Sen_2023}.
Therefore, spins in the two QDs live on Bloch spheres with axes (mis)aligned along their internal fields, and can only \textit{see} each other through the \textit{warped} lens of a spin precession (i.e., the two Bloch spheres are \textit{connected} by the rotation $W$).

Undoubtedly, the reader will be reminded of the previous discussion of polarimetry, only raised one dimension higher. This is understood by the fact that, unlike a \textit{flat} polarizer that rotates around a fixed axis, \textit{g}-tensors are \textit{three-dimensional}, and one may orient $\vec{B}$ in any direction in 3D space.

The analogy with polarimetry is particularly suited to describe even-parity inter-dot transitions (i.e., $(1,1)\leftrightarrow(0,2)$ charge occupation), owing to the Pauli exclusion principle. 
Spins occupying separate QDs will generally align with the internal fields and will therefore rotate as we modify the external $\vec{B}$. 
Conversely, when two spins occupy the same QD, the Pauli principle requires that they be in a singlet state ($\szt = \left(\ket{\uparrow\downarrow}-\ket{\downarrow\uparrow}\right)/\sqrt{2}$), which is a scalar under rotations, thus providing a \textit{fixed reference} onto which to project.
In keeping with the traditional polarimetry nomenclature, we shall therefore refer to the QD that may host two spins as the (fixed) \textit{analyzer} (red), and to the other as the (changeable) \textit{polarizer} (blue).

\section*{A Polarimeter for Spins} 

A general treatment of the $(1,1)\leftrightarrow(0,2)$ transition requires at least five states: the four spin orientations in the $(1,1)$ occupation and $\szt$ (see Fig.~\ref{fig1:polarimeter}(c) and \ref{SM:Hamiltonian}).
However, in the case of a sufficiently high magnetic field---which we shall discuss exclusively in this work---the lowest energy state consists of the two spins anti-aligned with their respective internal fields ($\dpda$). 
Thereby, the low-energy dynamics can be simplified by considering the effective two-level Hamiltonian
\begin{equation}
    H = \frac{1}{2}\begin{bmatrix}
        \varepsilon - \mu_{\rm B} \left(|\polcol{\gten_{\rm P} \vec{B}}| + |\ancol{\gten_{\rm A} \vec{B}}|\right) &~& \sqrt{2} \Lambda \tilde{t}_{\rm c}\\
        \sqrt{2} \Lambda^* \tilde{t}_{\rm c}  &~& - \varepsilon
    \end{bmatrix},
    \label{eq:H_ground}
\end{equation}
\noindent
where $\varepsilon$ is the DQD energy detuning, $\tilde{t}_{\rm c}= \sqrt{t_0^2 + |\vec{t}|^2}$ is the total (zero-field) tunnel coupling, and 
\begin{equation}
    |\Lambda| = |\braket{ \ancol{\uparrow_{\rm A}} | W | \polcol{\downarrow_{\rm P}}}| = \sin{\frac{\theta_{\rm SM}}{2}}
    \label{eq:lambda}
\end{equation}
\noindent
quantifies the \textit{spin misalignment} of the ($\vec{t}$-rotated) polarizer state over the analyzer. Particularly, $|\Lambda|^2$ represents the probability of a spin flip upon projection of the rotated image of the polarizer onto the analyzer's quantization axis. 
The complex spin dynamics can thus be summarized in the parameter $|\Lambda|$ (or $\theta_{\rm SM}$), which is manifest in the modulation of the magnitude of the high-field avoided crossing between the $(1,1)$ and $(0,2)$ ground states.
A more direct physical understanding can be gained by observing that $\theta_{\rm SM}$ can be defined through 
\begin{equation}
    \cos{\theta_{\rm SM}} = \frac{\left(\mathbf{R} \polcol{\gten_{\rm P} \vec{B}}\right)\cdot \left(\ancol{\gten_{\rm A} \vec{B}}\right)}{|\polcol{\gten_{\rm P} \vec{B}}||\ancol{\gten_{\rm A} \vec{B}}|} ,
    \label{eq:cos_Ts}
\end{equation}
\noindent
where $\mathbf{R}$ is the orthogonal matrix that represents the same rotation as $W$ in 3D space \cite{Sen_2023,Danon_Nazarov_2009}.
Equation~(\ref{eq:cos_Ts}) provides a geometrical interpretation of $\theta_{\rm SM}$ as the angle of misalignment between the internal field of the analyzer and the image of the polarizer as seen through the active medium.
Any reader well-versed in Lie algebras will not have missed the factor-of-two difference between the definitions of $\theta_{\rm SM}$ in $SO(3)$ (Eq.~(\ref{eq:cos_Ts})) and its double-covering $SU(2)$ (Eq.~(\ref{eq:lambda})), manifesting the fact that antiparallel spins are \textit{orthogonal} on the Bloch sphere\footnote{Paraphrasing Sir R. Penrose describing Dirac's famed belt trick, when it comes to spins, a $2 \pi$ rotation turns the world upside-down \cite{Penrose_rindler_1984,Newman_1942,Staley_2010}.}.

We engineer our spin polarimeter in a solid-state system that is subject to strong SOC: a p-type single-gate silicon nanowire transistor. In particular, we form a DQD between an electrostatically defined QD underneath the gate of the transistor (polarizer) and a boron acceptor in the channel (analyzer) \cite{vonHorstig_2024}, see Fig.~\ref{fig2:phi}a. From a measurement perspective, a key observation from \eqr{eq:H_ground} is that changes in $\Lambda$ directly correspond to modifications of the gap at the $\dpda\leftrightarrow\szt$ anticrossing ($\Delta_{\rm SO} = \sqrt{2} |\Lambda| \tilde{t}_{\rm c}$), and thus to the transition's quantum capacitance. This parameter quantifies the system's ability to tunnel between the two coupled states, and is thus directly related to $\Delta_{\rm SO}$~\cite{Mizuta2017}.
We probe the quantum capacitance by the dispersive interaction with a high-$Q$ superconducting microwave resonator ($\omega_{\rm r}/2 \pi = 2.1$\,GHz) connected to the gate of the transistor. Specifically, changes in reflected power from the resonator reflect changes in the quantum capacitance, thus allowing a direct measure of $|\Lambda|$~\cite{Peri_2023}. 

For our study, we focus on a charge transition with nominal charge occupation $(1,1)\leftrightarrow(0,2)$, (Fig.~\ref{fig2:phi}b) measured in the high-field limit ($|\vec{B}| = 0.5$\,T). The choice to have the boron as the doubly occupied QD avoids concerns about orbital, valley and higher spin-number states at high magnetic fields~\cite{vonhorstig2024electrical,Lundberg_2024}. The presence of signal at the inter-dot charge transition (ICT) indicates that the system is free to tunnel between the two charge (and spin) states $\dpda\leftrightarrow\szt$. However, at a specific orientation of the magnetic field (the \textit{magic angle}) the signal vanishes as tunneling becomes forbidden by PSB (Fig.~\ref{fig2:phi}b inset). 

We study this PSB transition experimentally and theoretically in Fig.~\ref{fig2:phi}c,d by sweeping the in-plane magnetic-field orientation, $\phi$, which includes the direction where we have preemptively located PSB. The experiment represents the spin equivalent of rotating a polarimeter's polarizer to find the angle where light gets fully transmitted (an aligned light polarimeter is maximally \textit{bright} while for spins it is maximally \textit{dark} as the dispersive signal vanishes). As we rotate the field, we modify the spin misalignment, resulting in three distinct regimes: (i) lifting of PSB, (ii) resonant and (iii) PSB, whose energy diagrams along the respective green, blue and pink lines are represented in Fig.~\ref{fig2:phi}f-h. 

In the PSB-lifting regime (green line), the spins are misaligned, and the strong SOC in the system manifests as a single and broad peak centered at $\varepsilon = \mu_{\rm B} (|\polcol{\gten_{\rm P} \vec{B}}| + |\ancol{\gten_{\rm A} \vec{B}}|)/2$ \cite{Mizuta2017,Esterli_Otxoa_Gonzalez-Zalba_2019,Peri_2023,vonhorstig2024electrical}. As we reduce $\Delta_{\rm SO}$, the peak increases in magnitude and sharpens, until $\Delta_{\rm SO}$ reaches the resonator frequency. Past that point, we reach the resonant regime (blue line), when $\Delta_{\rm SO} \leq \hbar \omega_{\rm r}$. In this intermediate region, we observe the signature of the resonant interaction between the DQD and the photon cavity \cite{Samkharadzeeaar2018,Toida_Nakajima_Komiyama_2013,Blais_Grimsmo_Girvin_Wallraff_2021}, which can be seen as incoherent spin rotations driven by the resonator~\cite{Mi_2018, Yu_2023, Crippa2019}. Finally, as the system approaches the \textit{magic angle} (pink line), PSB is recovered and the dispersive signal vanishes as the image of the polarizer through the SOC medium perfectly aligns with the analyzer ($\theta_{\rm SM} = 0$). In this direction, the $\dpda$-$\szt$ crossing is no longer avoided and spin and charge degrees of freedom remain uncoupled, leading to a vanishing dispersive signal and the loss of ability to drive the spins via the resonator. The electrical response of the system as the probe signal goes from adiabatic (green dashed line) to resonant (blue dashed line) can be described via our unified linear response theory, detailed in Ref.~\cite{Peri_2023} (see Fig.~\ref{fig2:phi}c and \ref{SM:RF}).

\begin{figure*}[!hbt]
    \centering
    \includegraphics[width = \textwidth]{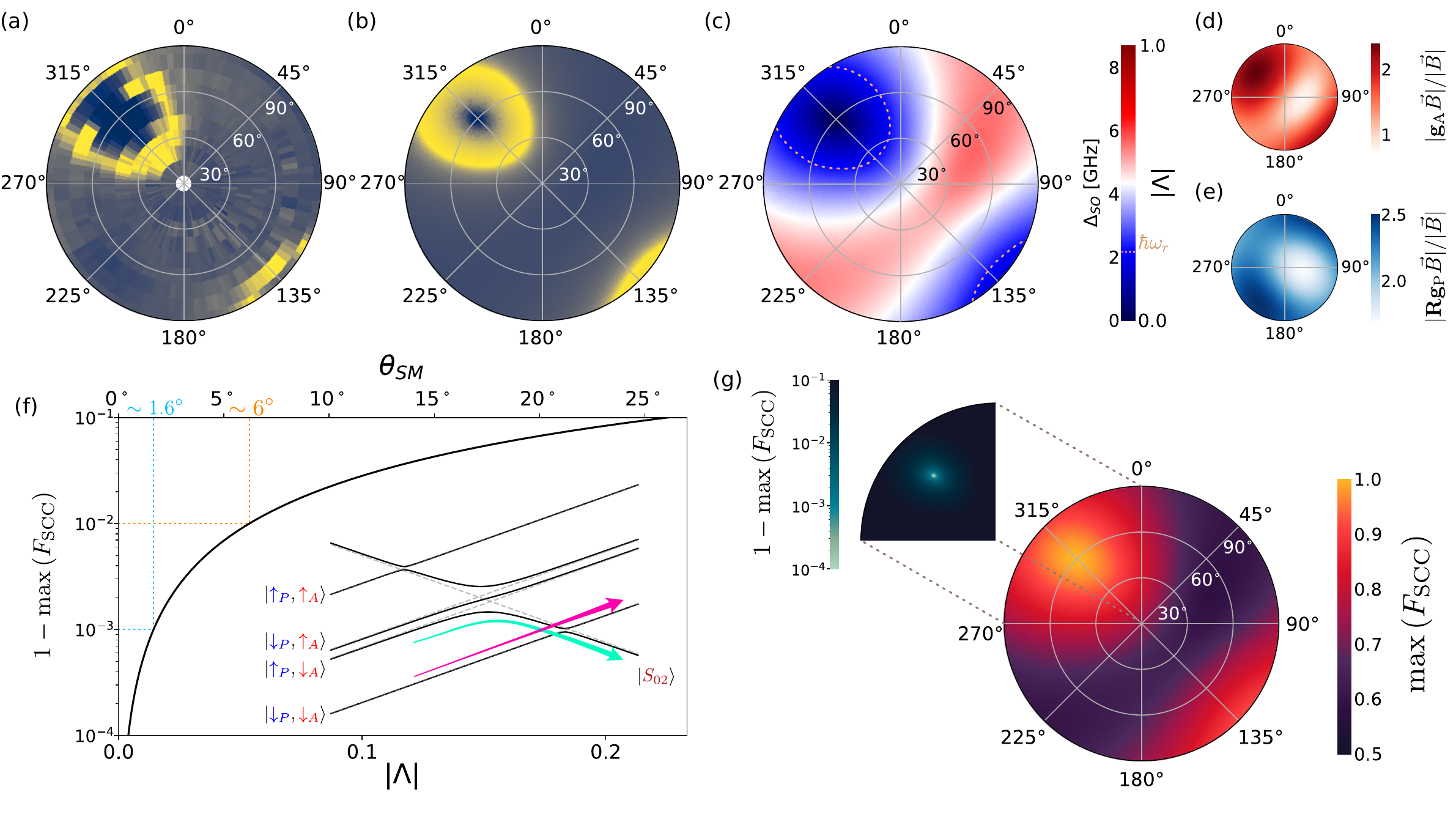}
    \caption{\textbf{Spin misalignment and spin readout.} Maximum dispersive signal data (a) and theory (b) of the charge transition for varying magnetic-field orientation, and (c) fitted spin-misalignment parameter $|\Lambda|$ and spin-orbit gap ($\Delta_{\rm SO} = \sqrt{2} \tilde{t}_{\rm c} |\Lambda|$). (d,e) Measured \textit{g}-tensors of the two quantum dots (see \ref{SM:Params}). (f) Spin-to-charge conversion (SCC) infidelity $1-F_{\textnormal{SCC}}$ (tunneling process in the inset) as a function of spin misalignment. (g) Expected $F_{\textnormal{SCC}}$ based on the parameters extracted from the data. The inset shows the minimum SCC infidelity on a logarithmic scale around the PSB direction.}
    \label{fig3:halo}
\end{figure*}

\section*{Polarimetry in the Third Dimension} 

Unlike traditional light polarimetry, polarimetry with spins is intrinsically three-dimensional in nature, thus requiring characterization of the system as a function of both azimuth and zenith. Thanks to time-reversal symmetry, the response is guaranteed to be equal for inversion of the field ($\vec{B} \rightarrow - \vec{B}$) \cite{Danon_Nazarov_2009,Mutter_Burkard_2021}. Therefore, it is sufficient to study only a solid angle of $2 \pi$. 
We do so exploiting our earlier observation that the dispersive signal can be used as a proxy for spin misalignment. Namely, the maximum signal will be small but finite for large $\Lambda$, peaking at the resonance condition $\Delta_{\rm SO} = \hbar \omega_{\rm r}$ and then vanishing as the system's dipole vanishes when the crossing is no longer avoided. 
In Fig~\ref{fig3:halo}a, we show that this trend appears as we approach the blockade from any direction, giving rise to a distinct ring-shaped region of intense signal encircling the PSB direction, which we name \textit{PSB halo}. This halo occurs at a zenith angle $\theta_{\rm SM} = 2\arcsin{\left(\hbar \omega_{\rm r} / \sqrt{2}\tilde{t}_{\rm c}\right)}$, describing a circle in the space of internal fields, and is then \textit{warped} in Cartesian coordinates by \textit{g}-tensor anisotropies.

To explore this further, we separately characterize the \textit{g}-tensors of the two QDs (Fig~\ref{fig3:halo}d,e, see \ref{SM:Params}). Firstly, we note that the PSB direction closely corresponds to the maximum Zeeman shift in the analyzer. This is a direct consequence of $\ancol{\gten_{\rm A}}$ having a stronger asymmetry than $\mathbf{R}\polcol{\gten_{\rm P}}$.
Similarly, we see the PSB halo \textit{squeezed} longitudinally. This is caused by both tensors having a larger gradient along the equator, resulting in the spins becoming misaligned faster in one direction.
\textit{En passant}, we stress that none of the principal axes of the \textit{g}-tensors are aligned with the nanowire, nor do the \textit{g}-tensors themselves respect the symmetry properties that one would have expected from the origin of the hole confinement (cylindrical for the QD and spherical for the boron).
This fact has already been observed in the literature \cite{Wang_2018,Crippa_2018,Piot2022,Huang_2024}, and it is testament to how SOC is a strong function of the local electromagnetic environment of each QD, highlighting the need for an accurate three-dimensional characterization without \textit{a priori} assumptions on symmetry \cite{Watzinger_Katsaros_2018,Liles_2024,Hendrickx_2024,Russell_2023}.
To this end, the PSB halo provides a clear signature in the hunt for the blockade, which differentiates between the disappearance of the dispersive signal due to PSB, modifications of the resonator impedance due to the magnetic field strength and/or angle~\cite{vonHorstig_2024}, and vanishing quantum capacitance as $\Delta_{\rm SO}$ grows too large~\cite{Han_2023}.

Lastly, we answer a pressing question in the literature \cite{Sen_2023,Han_2023}: does every system with strong SOC posses a \textit{magic angle}? Through the language of polarimetry, we can trivially conclude that the answer is \textit{no}, by noting that \eqr{eq:cos_Ts} describes a (three-dimensional) generalized eigenvalue problem.
Therefore, we are guaranteed that there will be at least one field orientation where $\mathbf{R} \polcol{\gten_{\rm P} \vec{B}} \parallel \ancol{\gten_{\rm A} \vec{B}}$, but the two quantization directions can be parallel ($\theta_{\rm SM}=0$) or antiparallel ($\theta_{\rm SM}=\pi$).
In the first case, we have normal PSB, while the second corresponds to an \textit{inverted} spin blockade, where the transition to $\szt$ is Pauli-blockaded for the \textit{anti}parallel $(1,1)$ states ($\upda$ and $\dpua$), as opposed to parallel states for the non-inverted case, and $\Delta_{\rm SO}$ is maximized \cite{Hendrickx_2021,Sen_2023}. This is in stark contrast with (two-dimensional) light polarimetry, which always presents a fully bright and a fully dark orientation.

\section*{Spin Misalignment and Spin Readout} 

Finally, we exploit the polarimetry analogy to investigate the consequences of spin misalignment on the readout of spin qubits.
Even-parity transitions are particularly interesting for quantum information processing. To this end, state manipulation may be performed in the (effective) $(1,1)$ occupation. Readout is then achieved by attempting to pulse to the $(0, 2)$ charge state. PSB will allow the (charge) transition to happen only if the two spins are in antiparallel states, offering a mechanism for spin-to-charge conversion (SCC) \cite{Oakes_2023,Huang_2024,Takeda_2024}. 
This process, however, requires states to safely navigate the energy manifold, traversing the singlet anticrossing adiabatically and $\Delta_{\rm SO}$ diabatically via a Landau-Zener (LZ) transition (Fig.~\ref{fig3:halo}f inset). 

SCC is conceptually akin to detecting a photon transmitted through a polarimeter, which highlights the possibility of conversion errors due to imperfect spin alignment. To quantify this effect, we introduce the concept of SCC fidelity ($F_{\textnormal{SCC}}\leq1$), a correction factor for any PSB-based spin readout, which must be taken into account when considering total readout fidelity, as for thermal effects in Elzermann readout~\cite{Keith_2019,Keith_2022}. 
In particular, $F_{\textnormal{SCC}}$ sets the fundamental upper limit for the spin-readout fidelity achievable via PSB.
Unlike photons (whose speed is constant), $F_{\textnormal{SCC}}$ strongly depends on the rate of the transition, and its analysis is further complicated by PSB being a non-demolition operation \cite{Huang_2024}. A full derivation is presented in the Supplementary Materials. Nevertheless, for small spin misalignment ($\theta_{\rm SM}\lesssim10^\circ$), the analogy with a polarimeter also holds quantitatively, as it may be shown that $\max \left(F_{\textnormal{SCC}}\right) \approx 1-(\theta_{\rm SM}/2)^2(1/2-\log|\theta_{\rm SM}/2|)$, i.e., the SCC infidelity depends on the square of the misalignment angle, multiplied by a logarithmic correction originating from the LZ formula.
Strikingly, this indicates that SCC errors are \textit{inevitable} unless the spins are perfectly aligned ($\theta_{\rm SM}=0$). Quantitatively, $F_{\textnormal{SCC}} > 99\%$ requires $\theta_{\rm SM}\lesssim6^\circ$, while $F_{\textnormal{SCC}} > 99.9\%$ requires $\theta_{\rm SM}\lesssim1.6^\circ$, and each additional $9$ needs \textit{exponentially} better alignment.
We stress that the above is an \textit{upper limit}, which requires pulsing at the optimal ramp rate $\dot{\varepsilon} \sim \tilde{t}_{\rm c}^2/\hbar$ (see \ref{SM:LZ}). Depending on the system, this may necessitate either pulses beyond the state of the art of signal generation \cite{vonhorstig2024electrical} or entail schemes unacceptably slow considering the ever-increasing requirement of fast spin readout \cite{Oakes_2023}.

The limits above are \textit{universal}, as $\theta_{\rm SM}$ is defined in the frame of reference of the internal fields. A conversion to the Cartesian frame of the applied external field requires knowledge of the \textit{g}-tensors of the system. For a concrete example, $F_{\textnormal{SCC}} > 99.9\%$ with the parameters found in this work requires magnetic-field accuracy below $1^\circ$ (Fig.~\ref{fig3:halo}g), and would require pulsing in excess of $30$\,MeV/s \cite{vonhorstig2024electrical}.
Notably, $F_{\textnormal{SCC}}$ may be of concern also for systems where SOC is weak (e.g., electrons in silicon). Spin misalignment due to small differences in \textit{g}-factors may be estimated $\theta_{\rm SM} \sim \delta g / \bar{g} $ ($\bar{g} \sim 2$ being the average \textit{g}-factor). Therefore, variations of $\delta g \sim 10^{-2}$, commonly observed in the literature \cite{Ruskov_Veldhorst_2018, Cifuentes_2024}, may make it challenging to reliably achieve fidelities above 99.9\%.

\section*{Outlook}

In its seminal \textit{Transactions of the Cambridge Philosophical Society}, G. G. Stokes prefaces his analysis of the dynamics of polarized light by stating that ``\textit{the object of the philosopher is not to complicate, but to simplify and analyze, so as to reduce phenomena to laws, which in their turn may be made the stepping-stones for ascending to a general theory which shall embrace them all}''\footnote{\textit{Verbatim}, G.G. Stokes, Ref.~\cite{Stokes_1851}}.
In this work, we believe we have upheld the spirit of this statement, for we have reduced the complexity of SOC in spin qubits to a simple law, made familiar by drawing parallels with the ubiquitous technique of polarimetry. 

From the analogy with polarimetry naturally arises the definition of the spin misalignment angle, which uniquely determines the avoided crossings in the Hamiltonian of a spin system in the presence of SOC. Importantly for quantum information processing, spin misalignment poses a fundamental upper limit for spin-to-charge conversion via PSB, which forms the basis of many spin-readout schemes. 
Thus, we have introduced the concept of SCC fidelity, which directly links the spin alignment to the achievable fidelity of spin readout.
This fundamental limit provides stringent requirements for the accurate and high-resolution characterization of the \textit{g}-tensors and tunnel couplings, highlighting the necessity of precisely identifying the \textit{magic} PSB direction (typically within less than $1^\circ$) as a prerequisite to achieving high-fidelity spin readout.

Moreover, a further challenge arises for the scalability of strongly spin-orbit-coupled systems. Particularly, SOC is dependent in the local environment of the spin sites, giving rise to sample-to-sample and device-to-device variability in \textit{g}-tensors~\cite{Martinez2022}. This translates into strong variability of the spin misalignment and in the \textit{magic} direction of the blockade (if this direction even exists), which may drastically differ even between neighboring QD pairs. 
Therefore, the limitation of being able to apply \textit{one} global external field may preclude the possibility to perform spin readout on all qubits above the fault-tolerance threshold, putting quantum engineers in the difficult position of deciding between poor-fidelity readout of many QDs or high-quality readout of only a selected few \textit{hero} qubits. 
We stress that this limit is fundamental in nature, and may only be circumvented if nanofabrication techniques become able to guarantee properties of nominally identical devices down to the defects and local strain in nanostructures.
Alternatively, the field may need to advance beyond PSB-based charge sensing, adopting methods such as \textit{in situ} dispersive readout, which can differentiate between spin states even when spin blockade is lifted.~\cite{vonhorstig2024electrical}.


\section*{Declarations}

\subsection*{Acknowledgements}
This research was supported by the European Union's Horizon 2020 research and innovation programme under grant agreement no.~951852 (QLSI), and by the UK's Engineering and Physical Sciences Research Council (EPSRC) via the Cambridge NanoDTC (EP/L015978/1). 
F.E.v.H. acknowledges funding from the Gates Cambridge fellowship (Grant No. OPP1144). 
M.F.G.Z. acknowledges a UKRI Future Leaders Fellowship [MR/V023284/1]. 
L.P. acknowledges the Winton Programme for the Physics of Sustainability for funding.

\subsection*{Competing Interests}

The Authors declare no competing financial or non-financial interests.

\subsection*{Data Avaliability}

The data that support the plots within this article and other findings of this study are available from the corresponding authors upon reasonable request.

\subsection*{Code Availability}

The unified linear response theory used for the simulations in this work is discussed in Ref.~\cite{Peri_2023}. The mathematical equations necessary to perform the simulations and data analysis are discussed in the Supplementary Materials.

\section*{Supplementary Materials}

\subsection{Hamiltonian of a Spin-Orbit-Coupled Double Quantum Dot}
\label{SM:Hamiltonian}

The Hamiltonian of a DQD is generally made of three components: (i) detuning ($H_\varepsilon$), (ii) Zeeman ($H_{\rm Z}$), and (iii) tunneling ($H_{\rm t}$).
In the presence of SOC, it is characterized by 17 parameters: the DQD detuning $\varepsilon$, the \textit{scalar} spin-conserving $t_0$ and spin-flip tunnel coupling vector $\vec{t}=(t_x, t_y, t_z)^{\rm T}$, and 6 parameters for each $3 \times 3$ \textit{g}-tensors. The requirement of time-reversal symmetry forces $t_0$ and $\vec{t}$ to be real, and the \textit{g}-tensors to be real and symmetric.

To more deeply understand the case of an even transition, as described in this work, we quickly discuss the case of an odd transition ($(1,0)\leftrightarrow (0, 1)$), to best highlight the physical origin of the spin-misalignment parameter $\Lambda$. 
In the literature, spin systems are traditionally modelled by choosing an arbitrary quantization axis (usually $\hat{z}$), causing the Hamiltonian to be highly dense \cite{Sen_2023,Danon_Nazarov_2009} as it convolves the rotations due to the \textit{g}-tensors and $\vec{t}$. In this work, instead, we choose the more natural basis where the quantization axis of each spin is aligned with the relative internal field, thus making the Zeeman contribution diagonal. In this basis the Hamiltonian reads 
\begin{align}
    H^{\textrm{odd}} =& H_\varepsilon^{\textrm{odd}} + H_{\rm Z}^{\textrm{odd}} + H_{\rm t}^{\textrm{odd}} \\
    H_\varepsilon^{\textrm{odd}} =& \frac{\varepsilon}{2} \left( \ket{\polcol{\uparrow_{\rm P}}}\bra{\polcol{\uparrow_{\rm P}}} + \ket{\polcol{\downarrow_{\rm P}}}\bra{\polcol{\downarrow_{\rm P}}}\right)\\
     - &\frac{\varepsilon}{2} \left(\ket{\ancol{\uparrow_{\rm A}}}\bra{\ancol{\uparrow_{\rm A}}} + \ket{\ancol{\downarrow_{\rm A}}}\bra{\ancol{\downarrow_{\rm A}}}\right) \nonumber \\
     H_{\rm Z}^{\textrm{odd}} =& \frac{\mu_{\rm B} }{2} \left| \polcol{\gten_{\rm P} \vec{B}}\right| \left( \ket{\polcol{\uparrow_{\rm P}}}\bra{\polcol{\uparrow_{\rm P}}} - \ket{\polcol{\downarrow_{\rm P}}}\bra{\polcol{\downarrow_{\rm P}}}\right)\\
     + &\frac{\mu_{\rm B} }{2} \left| \ancol{\gten_{\rm A} \vec{B}}\right| \left(\ket{\ancol{\uparrow_{\rm A}}}\bra{\ancol{\uparrow_{\rm A}}} - \ket{\ancol{\downarrow_{\rm A}}}\bra{\ancol{\downarrow_{\rm A}}}\right) \nonumber \\
     H_{\rm t}^{\textrm{odd}}=& \tilde{t}_{\rm c} \big( Q \ket{\polcol{\downarrow_{\rm P}}}\bra{\ancol{\downarrow_{\rm A}}} + Q^* \ket{\polcol{\uparrow_{\rm P}}}\bra{\ancol{\uparrow_{\rm A}}} \label{eq:app_H_odd}\\
      &~+ \Lambda \ket{\polcol{\downarrow_{\rm P}}} \bra{\ancol{\uparrow_{\rm A}}}  - \Lambda^* \ket{\polcol{\uparrow_{\rm P}}} \bra{\ancol{\downarrow_{\rm A}}}\big) + h.c. \nonumber
\end{align}
\noindent
where we point out that in odd transitions the choice of analyzer and polarizer is arbitrary. Above, we have defined
\begin{align}
    |Q| &= \left|\braket{ \ancol{\downarrow_{\rm A}} \left| W \right| \polcol{\downarrow_{\rm P}}}\right| = \cos{\frac{\theta_{\rm SM}}{2}}\label{eq:app_def_Q}\\
    |\Lambda| &= \left|\braket{\ancol{\uparrow_{\rm A}} \left| W \right| \polcol{\downarrow_{\rm P}}}\right| = \sin{\frac{\theta_{\rm SM}}{2}} ,\label{eq:app_def_Lambda}
\end{align}
\noindent
which highlights the physical interpretation of $\theta_{\rm SM}$ as the zenith angle of the image of the polarizer ($W \ket{\polcol{\downarrow_{\rm P}}}$) on a Bloch sphere, the axis of which is aligned with the internal field of the analyzer. 
The phases of $Q$ and $\Lambda$ depend on the choice of gauge for the two Kramers pairs, particularly on the (independent) choices of the $\hat{x}$ and $\hat{y}$ axes for the two separate Bloch spheres. In particular, it is always possible to find a gauge where both $Q$ and $\Lambda$ are real and positive.

It is interesting to point out how the literature traditionally refers to $\vec{t}$ as \textit{spin-flip} tunneling, as it causes a coupling between states of anti-aligned spins \cite{Danon_Nazarov_2009,Benito_2017,Yu_2023}. However, \eqr{eq:app_H_odd} shows how this is somewhat of a misnomer, as a finite $\Lambda$ (and thus spin-flip) may arise even if $\vec{t}=0$. This fact is promptly understood by noticing that if the two quantization axes are themselves misaligned, there will be a finite projection of the respective anti-aligned spins onto one another even in the absence of any rotation. This highlights the complexity of SOC, as its properties are a (highly nontrivial) combination of \textit{all} the Hamiltonian parameters.
In particular, from Eqs.~(\ref{eq:app_def_Q}) and (\ref{eq:app_def_Lambda}) we can see how $|Q|^2+|\Lambda|^2=1$, and their function in the Hamiltonian is to distribute the total (zero-field) tunnel coupling between spin-conserving ($Q$) and spin-flip ($\Lambda$) transitions. 
Hence, it would perhaps be more appropriate to apply the names spin-conserving and spin-flip parameters to $Q$ and $\Lambda$ rather than to $t_0$ and $\vec{t}$.

Strikingly, the complex behavior induced by the presence of SOC may be described by the same parameters $\Lambda$ and $Q$, which, perhaps even more surprisingly, may be employed to describe also even transitions. 
As mentioned, in this work we consider the $(1,1)\leftrightarrow (0, 2)$ transition, which is well-described using only five states: $\upua$, $\upda$, $\dpua$, and $\dpda$ in the $(1,1)$ region and $\szt$ in the $(0,2)$ occupation. We neglect any excited triplet arising from higher orbital, or valley states, which in our experimental setup is justified by the use of a boron atom as the (doubly occupied) analyzer. Moreover, as above, we discuss the basis in which spins are aligned with their respective internal field. In this basis, the Hamiltonian reads
\begin{align}
    H^{\textrm{even}} =& H_\varepsilon^{\textrm{even}} + H_{\rm Z}^{\textrm{even}} + H_{\rm t}^{\textrm{even}} \\
    H_\varepsilon^{\textrm{even}} =& \frac{\varepsilon}{2} \mathbb{I} - \varepsilon \ket{\singcol{S_{02}}}\bra{\singcol{S_{02}}}\\
     H_{\rm Z}^{\textrm{even}} =&  \\
     \frac{\mu_{\rm B} }{2}\bigg(&\big(|\polcol{\gten_{\rm P} \vec{B}}| + |\ancol{\gten_{\rm A} \vec{B}}|\big) \diagproj{\polcol{\uparrow_{\rm P}}, \ancol{\uparrow_{\rm A}}} \nonumber \\
    -&\big(|\polcol{\gten_{\rm P} \vec{B}}| + |\ancol{\gten_{\rm A} \vec{B}}|\big) \diagproj{\polcol{\downarrow_{\rm P}}, \ancol{\downarrow_{\rm A}}} \nonumber \\
    +&\big(|\polcol{\gten_{\rm P} \vec{B}}| - |\ancol{\gten_{\rm A} \vec{B}}|\big) \diagproj{\polcol{\uparrow_{\rm P}}, \ancol{\downarrow_{\rm A}}} \nonumber \\
    -&\big(|\polcol{\gten_{\rm P} \vec{B}}| - |\ancol{\gten_{\rm A} \vec{B}}|\big) \diagproj{\polcol{\downarrow_{\rm P}}, \ancol{\uparrow_{\rm A}}} \nonumber \bigg)\\
     H_{\rm t}^{\textrm{even}}=& \\
     \frac{\tilde{t}_{\rm c}}{\sqrt{2}} \big( &Q \dpua\bra{\singcol{S_{02}}}  - Q^* \upda\bra{\singcol{S_{02}}}\\
     +& \Lambda \dpda\bra{\singcol{S_{02}}} + \Lambda^* \upua\bra{\singcol{S_{02}}}
      \big) + h.c. , \nonumber
\end{align}
\noindent
where $\mathbb{I}$ is the ($5 \times 5$) identity. Notably, we see how, as for the odd case, both $H_\varepsilon$ and $H_{\rm Z}$ are diagonal in this basis, while $H_{\rm t}$ connects the $(0,2)$ singlet with the $(1,1)$ states. In particular, recalling the definition of $\szt$, we can see how also in this case $Q$ describes transitions where the transitioning spin's alignment is conserved, while $\Lambda$ describes transition where spin-flip occurs.

\begin{figure}[htb!]
    \centering
    \includegraphics[width = 0.9\linewidth]{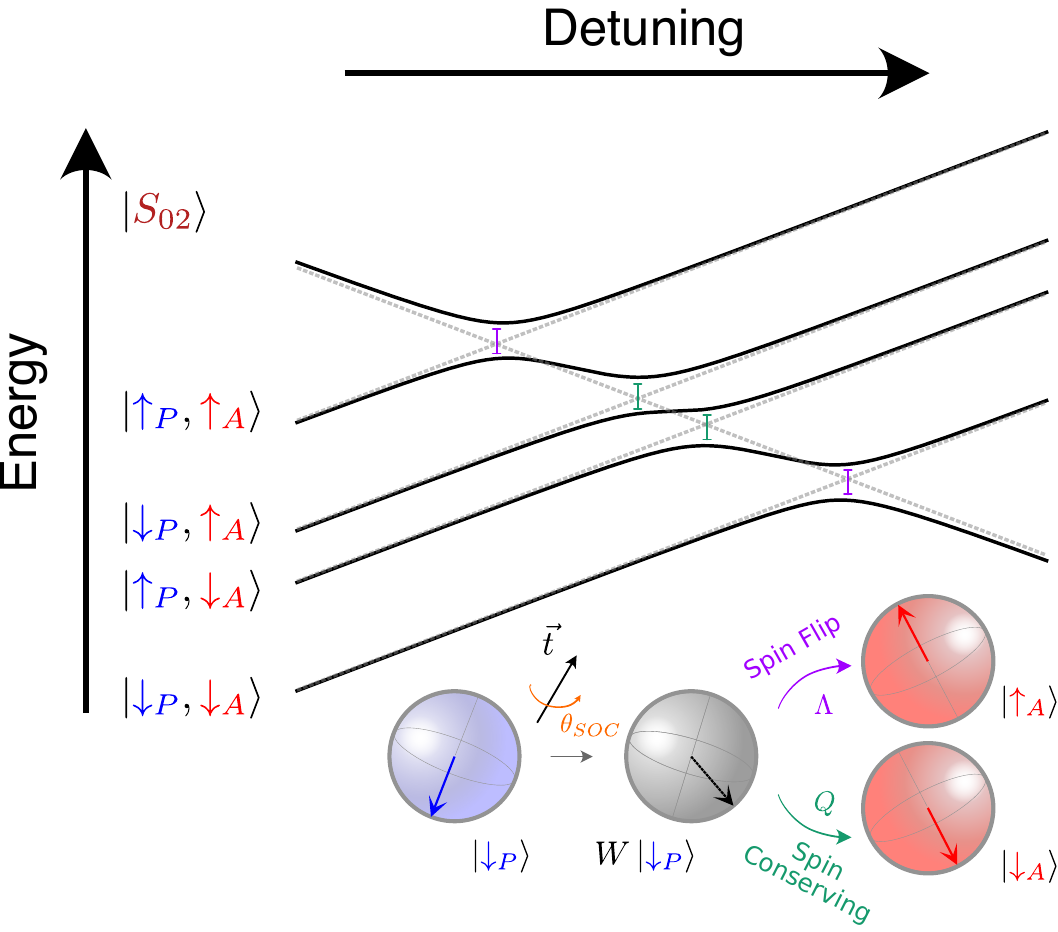}
    \caption{\textbf{Energy diagram of a spin-orbit-coupled even-parity transition.} The colors emphasize the effect of $Q$ and $\Lambda$ on the avoided crossing due to spin-flip (purple) and spin-conserving (green) tunneling, pictorially shown in the inset.}
    \label{figapp:delta}
\end{figure}

\subsection{Dispersive Sensing of Lifted Blockades}
\label{SM:RF}

Performing polarimetry with spins requires measuring the dispersive signal of an anticrossing that may vanish (at the PSB direction). Thus, interpretation of the experimental data requires a deep understanding of the dispersive interaction between a resonator and a quantum system. 
Electrically, the quantum system is effectively in parallel (Fig.~\ref{fig2:phi}) with the superconducting resonator. If we assume, as is the case in this work, that the radio-frequency excitation through which the system is probed is small enough that the system responds \textit{linearly}, the DQD's behavior may be summarized by a (complex) admittance $Y$ \cite{Peri2024beyondadiabatic,oakes2022quantum}. If we also take, as we shall for simplicity, that the coherent response dominates over the Sisyphus and Hermes components, the admittance reads \cite{Peri_2023}
\begin{equation}
    Y(\varepsilon) = \frac{\alpha^2e^2}{\hbar} \sum_{m, n} \frac{ (p_m-p_n) \left| \braket{\phi_m|\Pi|\phi_n} \right|^2}{\omega_{\rm r} - (E_n-E_m) - i \gamma},
    \label{eq:app_Y_gen}
\end{equation}
\noindent
where $\alpha$ is the (differential) DQD lever arm, $\ket{\phi_m}$ and $E_m$ are the eigenstate and eigenenergy of the Hamiltonian (as a function of $\varepsilon$) occupied with probability $p_m$, $\Pi = {\rm d}H/{\rm d}\varepsilon$ is the dipole operator, and $\gamma$ is the decoherence rate. 
If we assume, as we have in the main text, that the external field is large enough that the DQD may be effectively considered as a two-level system (\eqr{eq:H_ground}), the admittance simplifies to \cite{Peri_2023}
\begin{equation}
    Y(\varepsilon) = i \frac{\alpha^2 e^2}{2 \hbar} \frac{\Delta_{\rm SO}^2}{\Delta E} \frac{\omega_{\rm r}}{\Delta E^2 + \left(\gamma + i \omega_{\rm r}\right)^2},
    \label{eq:app_Y_TLS}
\end{equation}
\noindent
where $\Delta E=\sqrt{\Delta_{\rm SO}^2 + \varepsilon^2}$ (we assume negligible probability in the excited state).
From this we obtain the dispersive signal as the reflection coefficient of a resonator with a variable impedance in parallel \cite{Ibberson_2021},
\begin{equation}
    \Gamma(\omega) \propto \frac{1}{ i(\omega - \omega_{\rm r}) + \kappa/2 + \eta Y}, 
\end{equation}
\noindent
where $\kappa$ is the bandwidth of the resonator and $\eta$ is a measure of the (coherent) qubit--resonator coupling \cite{Ruskov_dynlong_2024}. We note that, for small $\eta$, the changes in reflection coefficient due to changes in the quantum system's admittance (i.e., because of changes in detuning), can be approximated as 
\begin{equation}
    |\Delta \Gamma| \propto |Y|.
\end{equation}

\begin{figure}[htb!]
    \centering
    \includegraphics[width = 0.7\linewidth]{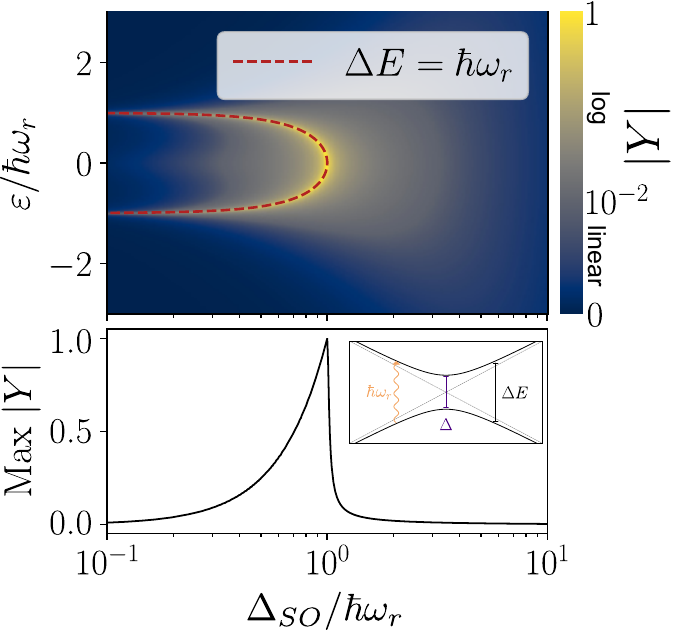}
    \caption{\textbf{Dispersive sensing of a lifted blockade.} Absolute value of the equivalent admittance of a two-level (avoided) crossing for varying energy gap ($\Delta_{\rm SO}$) probed at finite frequency $\omega_{\rm r}$.}
    \label{fig:app_delta}
\end{figure}

The behavior of \eqr{eq:app_Y_TLS} is shown in Fig.~\ref{fig:app_delta}, where we see the same trend observed in the main text when discussing Fig.~\ref{fig2:phi}. In particular, we see how the admittance, and thus the dispersive signal, increase with decreasing $\Delta_{\rm SO}$, and the maximum signal is reached when the anticrossing matches the resonator frequency. If the gap is larger than the photon energy, $|Y|$ takes the form of a single, zero-centered peak with sharply decreasing height proportional to $1/\Delta_{\rm SO}$. When, instead, $\Delta_{\rm SO} \leq \hbar \omega_{\rm r}$, the system responds strongly when the resonance condition is met (red dashed line), and the divergence in the admittance is cured by a finite decoherence rate $\gamma$. As the gap approaches zero (i.e., at the blockade), we see that the signal also vanishes. This is caused by the matrix element of the system's dipole in \eqr{eq:app_Y_gen}, which decreases when the curvature of the energy level diminishes, and vanishes exactly when the crossing is no longer avoided.

\subsection{Characterization of Physical Parameters} 
\label{SM:Params}

\begin{figure}[htb!]
    \centering
    \includegraphics[width = 0.9\linewidth]{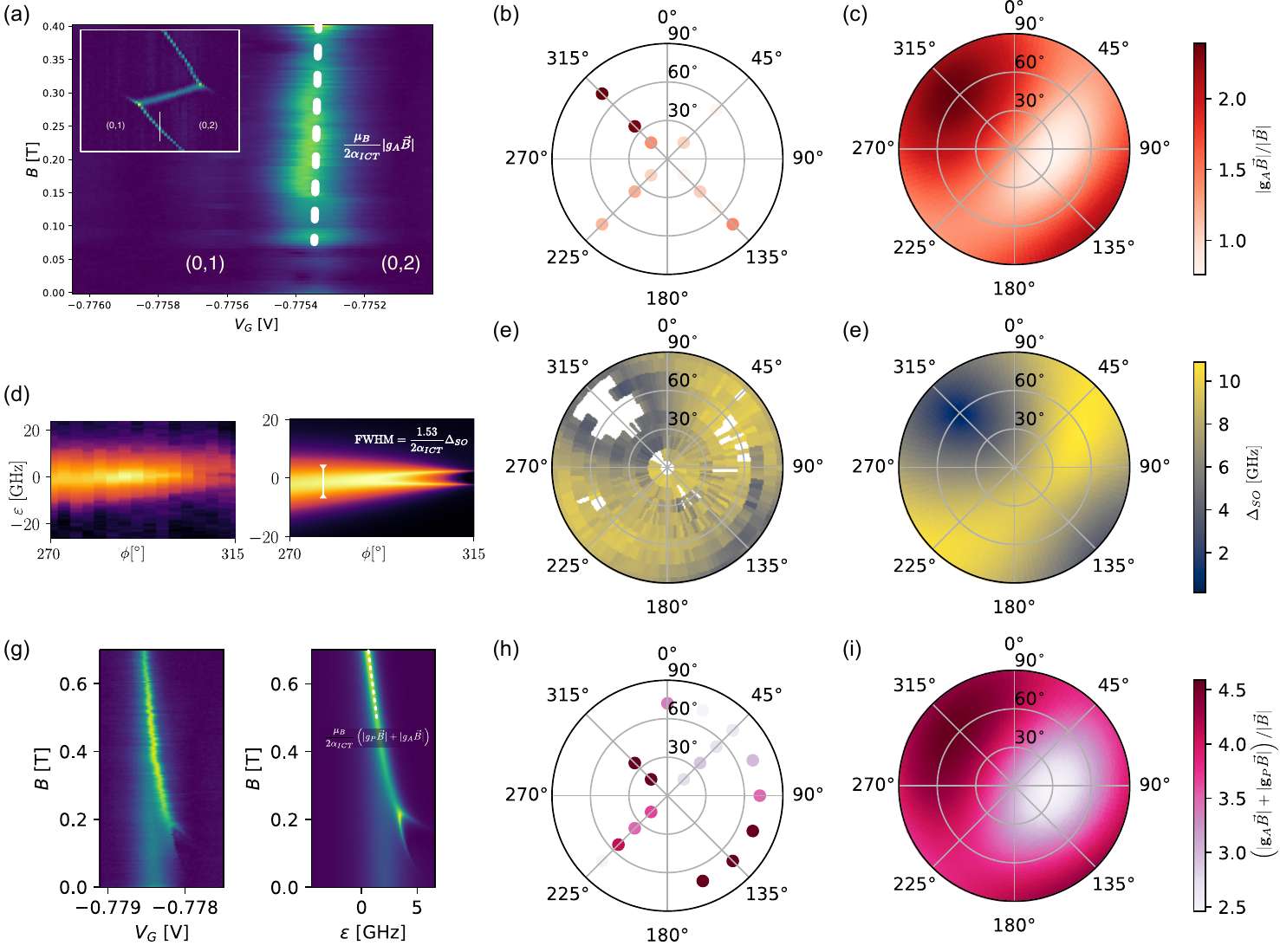}
    \caption{\textbf{\textit{g}-tensor and tunnel-coupling characterization} (a-c) Measurement of the \textit{g}-tensor of the boron acceptor (analyzer). The $(1,0)\leftrightarrow (0,2)$ charge transition occurs at $\varepsilon_{\rm BRT} = \mu_{\rm B} |\ancol{\gten_{\rm A}} \vec{B}|/2$, thus from its slope at various magnetic field directions (b) it is possible to reconstruct $\ancol{\gten_{\rm A}}$ (c). (d-f) Measurement of $\Delta_{\rm SO}$. The full-width half-maximum (FWHM) of the dispersive signal from the $(1,1)\leftrightarrow (0,2)$ is $1.53\Delta_{\rm SO}/2$, from which we fit $\Delta_{\rm SO}\propto |\Lambda|$. (g-i) To decouple the rotation due to $\vec{t}$ and $\polcol{\gten_{\rm P}}$, we measure the $(1,1)\leftrightarrow (0,2)$ transition as a function of magnetic field (g). At high field, the transition occurs at $\varepsilon_{\rm ICT} = \mu_{\rm B} (|\ancol{\gten_{\rm A}} \vec{B}| + |\polcol{\gten_{\rm P}} \vec{B}|)/2$. From its slope in various magnetic-field directions (h) it is possible to reconstruct $|\ancol{\gten_{\rm A}} \vec{B}| + |\polcol{\gten_{\rm P}} \vec{B}|$ (i).}
    \label{fig:SMData}
\end{figure}

In this work, we separately characterize the \textit{g}-tensors of both QDs, as well as the direction-dependent tunnel coupling, by making use of the visible boron-to-reservoir transition (BRT) adjacent to the inter-dot charge transition (ICT) discussed in the main text. When sweeping the magnetic-field angle, the BRT occurs at $\varepsilon_{\rm BRT} = \mu_{\rm B} |\ancol{\gten_{\rm A}} \vec{B}|/2$. Thus, from the slope at various magnetic-field orientations, it is possible to reconstruct $\ancol{\gten_{\rm A}}$ (Fig.~\ref{fig:SMData}a-c).

To characterize the ICT parameters, we study the ICT for varying field orientations (Fig.~\ref{fig:SMData}d-f). The full-width half-maximum (FWHM) of the dispersive peak is $1.53 \Delta_{\rm SO} \approx 2.2 |\Lambda| \tilde{t}_{\rm c}$ \cite{Esterli_Otxoa_Gonzalez-Zalba_2019,Peri_2023} (excluding the regions where $\Delta_{\rm SO} < \hbar \omega_{\rm r}$ and peak-splitting occurs). 
Fitting these data leads to accurate reconstruction of $|\Lambda|$, and hence we fit $\mathbf{R}\polcol{\gten_{\rm P}}$ (Fig.~\ref{fig3:halo}d,e), using the previous knowledge of $\ancol{\gten_{\rm A}}$.

To decouple the rotation due to $\vec{t}$ and $\polcol{\gten_{\rm P}}$, we measure the $(1,1)\leftrightarrow (0,2)$ transition as a function of magnetic field.
At high field, the ICT occurs at $\varepsilon_{\rm ICT} = \mu_{\rm B} (|\ancol{\gten_{\rm A}} \vec{B}| + |\polcol{\gten_{\rm P}} \vec{B}|)/2$. Thus, fitting this slope for various field directions, we derive $|\ancol{\gten_{\rm A}} \vec{B}| + |\polcol{\gten_{\rm P}} \vec{B}|$ (Fig.~\ref{fig:SMData}g-i).

The parameters obtained through this process are
\begin{align}
&\ancol{\gten_{\rm A}} = \begin{pmatrix}
    1.7 & -0.4& 0.3 \\
    -0.4 & 1.7 & -0.6 \\
    0.3 & -0.6 & 1.1
\end{pmatrix}\\
&\polcol{\gten_{\rm P}} = \begin{pmatrix}
    2.4& 0.1 &0.0\\
    0.1 & 2.2 & -0.3\\
    0.0 & -0.3 & 1.9
\end{pmatrix}\\
&\vec{t} = \begin{pmatrix}
    -4.8, 4.7 , -3.4
\end{pmatrix}^{\rm T} \,\textrm{GHz} \\
&t_0 = 9.7\,\textrm{GHz}.
\end{align}
Recording the BRT in the same sweep of the ICT has the added benefit of allowing for accurate calibration of the intensity of the dispersive signal accounting for changes in the superconducting resonator with magnetic field, as the BRT peak height is independent of field direction \cite{Peri2024beyondadiabatic}.

\subsection{Landau-Zener Transitions and Spin-to-Charge Conversion Fidelity}
\label{SM:LZ}

Readout of spin-qubit states through PSB (whether dispersively or via charge sensing) requires the states to safely traverse several avoided crossings of potentially different magnitudes. We discuss here the simplest case, where the computational basis consists of the ground ($\dpda$) and first excited ($\upda$) states, i.e., a spin qubit is the polarizer and the analyzer is an ancilla for readout. The results presented here can can, however, be generalized to any state.

\begin{figure}[htb!]
    \centering
    \includegraphics[width = 0.9\linewidth]{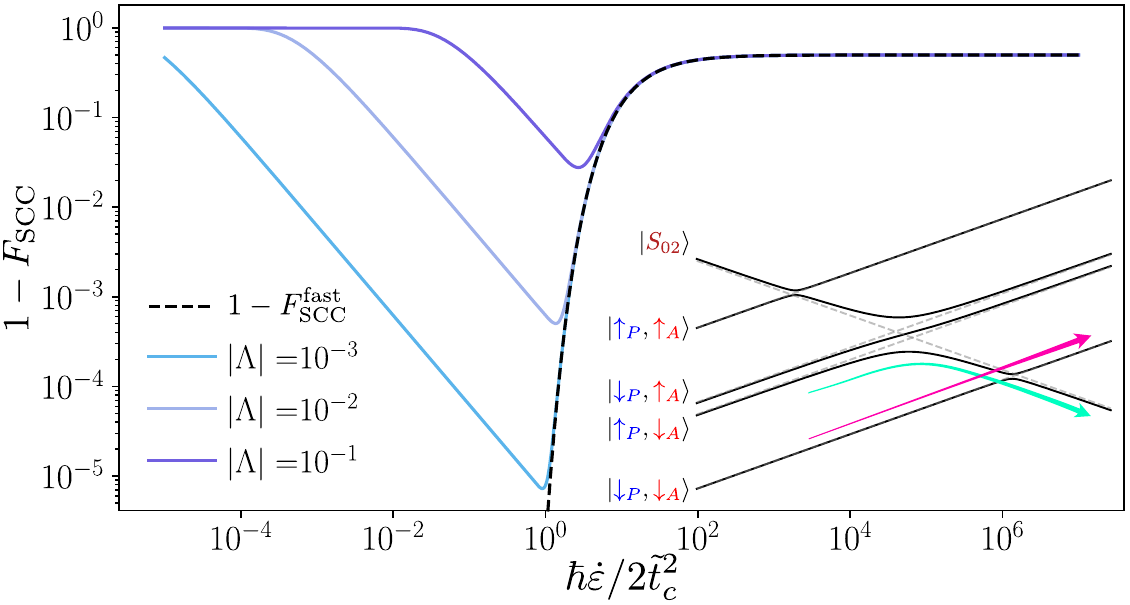}
    \caption{\textbf{Spin-to-charge conversion fidelity.} Fidelity of the spin-to-charge conversion process via PSB (inset) for different spin misalignment.}
    \label{fig:LZ_PSB}
\end{figure}

To perform charge readout of these states, one must move the states from the $(1,1)$ occupation where then computation occurs to the $(0,2)$ region, where SCC occurs, as (charge) tunneling of the $\dpda$ state is forbidden because of PSB. 
To successfully perform this operation, the parallel state must diabatically cross the spin-orbit anticrossing. This spin-flip process (from $T_-$ to $\szt$) occurs with a probability given by the Landau-Zener formula \cite{Glasbrenner_Schleich_2023,Shevchenko_2010}
\begin{equation}
    P_{sf} = 1-\exp\left(-4 \pi |\Lambda|^2 \frac{\tilde{t}_{\rm c}^2}{\hbar \dot{\varepsilon}}\right),
\end{equation}
\noindent
where, as in the main text, $\dot{\varepsilon}$ is the detuning ramp rate.
This is not, however, the only (coherent) error that can occur. The excited state, in fact, needs to adiabatically traverse the (triple) $\upda$-$\dpua$-$\szt$ crossing, while also diabatically avoiding a spin flip due to finite $\Delta_{\rm SO}$.
The first probability can be calculated exactly via the Demkov-Ostrovsky model, and reads \cite{Band_Avishai_2019,Chernyak_2020,Demkov_Ostrovsky_2001,Militello_2019}
\begin{equation}
    P_{ct} = 1-\exp\left(-8 \pi |Q|^2 \frac{\tilde{t}_{\rm c}^2}{\hbar \dot{\varepsilon}}\right),
\end{equation}
\noindent
which represents the probability of the charge transition.
We conclude that the fidelity of the SCC process is
\begin{equation}
    F_{\textnormal{SCC}} = \frac{\left(1-P_{sf}\right) \left(1+P_{ct}\right)}{2},
\end{equation}
\noindent
which, recalling the fact that $|\Lambda|^2 + |Q|^2=1$, can never reach unity unless $|\Lambda| = 0$ (Fig.~\ref{fig:LZ_PSB}). 
As a matter of fact, it can be shown that 
\begin{equation}
    \max\left(F_{\textnormal{SCC}}\right) = \frac{1}{1+\zeta/2}\left(\frac{\zeta}{1+\zeta/2}\right)^{\frac{\zeta}{2}} ,
    \label{eq:maxF}
\end{equation}
\noindent
where $\zeta = |\Lambda|^2/|Q|^2 = \tan^2(\theta_{\rm SM}/2)$. This is achieved for a ramp rate 
\begin{equation}
    \dot{\varepsilon}_{\max} = \frac{\tilde{t}_{\rm c}^2}{\hbar} \frac{8 \pi |Q|^2}{\log{\left(\frac{1}{|\Lambda|^2} - \frac{1}{2}\right)}}.
    \label{eq:maxRamp}
\end{equation}
In the case of quasi-alignment to the PSB direction ($|\Lambda|\ll1$), \eqr{eq:maxF} can be approximated as
\begin{equation}
    1- \max\left(F_{\textnormal{SCC}}\right) \approx |\Lambda|^2\left(\frac{1}{2}-\log |\Lambda|\right),
\end{equation}
\noindent
which, considering that for small spin-misalignment $|\Lambda|\approx\theta_{\rm SM}/2$, is equivalent to the expression in the main text.

We must point out, however, that, depending on the value of $\tilde{t}_{\rm c}$, obtaining the ramp rate in \eqr{eq:maxRamp} may require pulse rates too fast for the state-of-the-art of waveform generation \cite{vonhorstig2024electrical} or too slow to be acceptable considering the ever-increasing demand of fast spin readout, necessary for quantum error correction.
On the topic of ramp rates, unlike what may appear at first glance, faster pulses are not always beneficial. Rather, once the ramp rate exceeds $\dot{\varepsilon}_{\max}$, $F_{\textnormal{SCC}}$ drops very rapidly, following a universal limit (black dashed line in Fig.~\ref{fig:LZ_PSB})
\begin{equation}
    F^{\textnormal{fast}}_{\textnormal{SCC}} = 1-\frac{1}{2}\exp\left(-8 \pi \frac{\tilde{t}_{\rm c}^2}{\hbar \dot{\varepsilon}}\right),
\end{equation}
\noindent 
which arises from diabatic transitions at the spin-conserving anticrossing.

\subsection{Fabrication details} 

The transistor used in this study is the same as in Ref.~\cite{vonhorstig2024electrical} and consists of a single gate silicon-on-insulator (SOI) nanowire transistor with a channel width of 120~nm, a length of 60~nm and height of 8~nm. It was fabricated on an SOI substrate with a 145~nm-thick buried oxide and with a boron doping density of $5\cdot 10^{17}$~ cm$^{-3}$. The silicon layer was patterned to create the channel using optical lithography, followed by a resist trimming process. The transistor gate stack consists of 1.9 nm HfSiON capped by 5 nm TiN and 50 nm polycrystalline silicon, leading to a total equivalent oxide thickness of 1.3 nm. After gate etching, a SiN layer (10 nm) was deposited and etched to form a first spacer on the sidewalls of the gate, then 18-nm-thick Si raised source and drain contacts were selectively grown before source/drain extension implantation and activation annealing. A second spacer was formed, followed by source/drain implantations, an activation spike anneal and silicidation (NiPtSi). The nanowire quantum device and superconducting resonator were connected via on-chip aluminum bond wires.

\subsection{Measurement set-up} 

Measurements were performed at the base temperature of a dilution refrigerator (T $\sim$ 10\,mK). Low-frequency signals ($V_\text{g}$, $V_\text{bg}$) were applied through cryogenic filters, while radio-frequency readout tones were applied through filtered coaxial lines to a coupling capacitor connected to the RF resonator. The resonator consists of a NbTiN superconducting spiral inductor ($L \sim 30$\,nH), coupling capacitor ($C_{\rm c} \sim 40$\,fF) and low-pass filter fabricated by Star Cryoelectronics. For exact details, see Ref. \cite{vonHorstig_2024}. The PCB was made from 0.8-mm-thick RO4003C with an immersion silver finish. The reflected RF signal was amplified at 4\,K and room temperature, followed by quadrature demodulation (Polyphase Microwave AD0540B), from which the amplitude and phase of the reflected signal were obtained (homodyne detection). Magnetic fields were applied using a 5\,T/1\,T/1\,T American Magnetics vector magnet with the 5\,T direction aligned in the plane of the device chip.

%


\end{document}